\documentclass[sigconf,nonacm]{acmart}
\AtBeginDocument{%
  \providecommand\BibTeX{{%
    \normalfont B\kern-0.5em{\scshape i\kern-0.25em b}\kern-0.8em\TeX}}}

\usepackage{todonotes}
\usepackage{tikz}
\usetikzlibrary{calc,matrix}

\makeatletter
    \let\matamp=&
    \catcode`\&=13
    \def&{%
        \iftikz@is@matrix%
            \pgfmatrixnextcell%
        \else%
            \matamp%
        \fi%
    }
\makeatother

\newcounter{lines}
\def\endlr{\stepcounter{lines}\\}

\newcounter{vtml}
\newif\ifvtimelinetitle
\newif\ifvtimebottomline

\tikzset{
    description/.style={column 2/.append style={#1}},
    timeline color/.store in=\vtmlcolor,
    timeline color=red!80!black,
    timeline color st/.style={fill=\vtmlcolor,draw=\vtmlcolor},
    use timeline header/.is if=vtimelinetitle,
    use timeline header=false,
    add bottom line/.is if=vtimebottomline,
    add bottom line=false,
    timeline title/.store in=\vtimelinetitle,
    timeline title={},
    line offset/.store in=\lineoffset,
    line offset=4pt,
}

\NewEnviron{vtimeline}[1][]{%
    \setcounter{lines}{1}%
    \stepcounter{vtml}%
    \begin{tikzpicture}[column 1/.style={anchor=east},
        column 2/.style={anchor=west},
        text depth=0pt,text height=1ex,
        row sep=1ex,
        column sep=1em,
        #1
    ]
        \matrix(vtimeline\thevtml)[matrix of nodes]{\BODY};
        \pgfmathtruncatemacro\endmtx{\thelines-1}

        \path[timeline color st]
            ($(vtimeline\thevtml-1-1.north east)!0.5!(vtimeline\thevtml-1-2.north west)$)--
            ($(vtimeline\thevtml-\endmtx-1.south east)!0.5!(vtimeline\thevtml-\endmtx-2.south west)$);

        \foreach \x in {1,...,\endmtx}{
            \node[circle,timeline color st, inner sep=0.15pt, draw=white, thick]
            (vtimeline\thevtml-c-\x) at
            ($(vtimeline\thevtml-\x-1.east)!0.5!(vtimeline\thevtml-\x-2.west)$){};
                \draw[timeline color st](vtimeline\thevtml-c-\x.west)--++(-3pt,0);
        }

        \ifvtimelinetitle%
            \draw[timeline color st]([yshift=\lineoffset]vtimeline\thevtml.north west)--
                ([yshift=\lineoffset]vtimeline\thevtml.north east);

            \node[anchor=west,yshift=16pt,font=\large]
                at (vtimeline\thevtml-1-1.north west)
                {\textsc{Timeline \thevtml}: \textit{\vtimelinetitle}};
        \else%
            \relax%
        \fi%

        \ifvtimebottomline%
            \draw[timeline color st]([yshift=-\lineoffset]vtimeline\thevtml.south west)--
            ([yshift=-\lineoffset]vtimeline\thevtml.south east);
        \else%
            \relax%
        \fi%
    \end{tikzpicture}
}

\begin{document}

\title{Teaching Energy-Efficient Software -- An Experience Report}

\author{Henrik Bærbak Christensen}
\affiliation{%
  \institution{Computer Science, Aarhus University}
  \country{Denmark}}
\email{hbc@cs.au.dk}

\author{Maja Hanne Kirkeby}
\affiliation{%
  \institution{Roskilde University}
  \country{Denmark}}
\email{majaht@ruc.dk}

\author{Bent Thomsen}
\author{Lone Leth Thomsen}
\affiliation{%
  \institution{Department of Computer Science, Aalborg University}
  \country{Denmark}}
\email{bt,lone@cs.aau.dk}

\renewcommand{\shortauthors}{Christensen et al.}

\begin{abstract} 
  Environmental sustainability is a major and relevant challenge
  facing computing. Therefore, we must start teaching theory,
  techniques, and practices that both increase an awareness in our
  student population as well a provide concrete advice to be applied
  in practical software development. In this experience report, we
  focus on energy consumption of executing software, and describe
  teaching approaches from three different universities that all
  address software energy consumption in various ways. Our main
  contribution is reporting lessons learned from these experiences and
  sketching some issues that teachers must be aware of when designing
  learning goals, teaching material and exercises.
\end{abstract}




\maketitle

\section{Introduction}

Mankind is facing a huge challenge in the face of climate breakdown,
and all our endeavours must be scrutinized in order to reduce energy
consumption as well as switch to more sustainable sources.  Several
proposals for introducing sustainability in ICT educations have been
put forward
\cite{turkin2018software,mann2010computing,klimova2016international},
with \cite{pollock:2019,peters2023sustainability} providing
comprehensive overviews.
%
Our use of
computing is increasing and it is thus important that we also consider
the energy consumed by our software systems---and as teachers
consider how to increase awareness and knowledge about how software is
developed to minimize its energy footprint while maintaining adequate
levels of service. In a survey reported in \cite{saraiva2021bringing}, 94.7\%(18) of respondents answered yes, while only 5.3\%(1) answered no to the question do you 
“believe software engineering students should be taught energy efficient practices”. 
\cite{saraiva2021bringing} and \cite{oprescu2022energy} present suggestions for topics which could be included in teaching programs about software energy consumption.

In this paper, we present and discuss three different approaches to
teaching about software energy consumption, taught at three different
universities in (country). Section~\ref{section:ruc} describes an
elective course for master students whose focus is on energy
consumption; Section~\ref{section:saip} describes part of a broader
software architecture course focused on energy efficient architecture;
while Section~\ref{section:aau} describes a specialization at master
level in energy aware programming.

\section{Sustainability and Energy Consumption}

Environmental sustainability is a vast topic, and in the context of
computing often termed \emph{green information and communication
  technology (ICT)} or simply \emph{green computing}: activities that
minimise the negative impact of ICT on the environment and optimise
the positive impact ICT can have~\cite{verdecchia2017green}.  This
entails reducing the environmental impact of computing equipment and
software through making products recyclable, extending its lifetime,
use renewable energy resources in data centers, and maximizing energy
efficiency of software, etc.

As teachers in software engineering and computer science, our primary
focus is software architecture, and thus our primary perspective is
what can be done at the software level: Designing, developing and
operating software programs and architectures that minimize energy
consumption. In the SAiP course (Section~\ref{section:saip}) the focus
is by choice narrowly defined as a focus on the architectural quality
attribute~\cite{bass:2021} \emph{energy-efficiency}, defined as

\begin{quote}
  {\bfseries Energy-Efficiency} is concerned with  \emph{the
    system’s ability to minimize energy consumption while providing
    it’s services.}
\end{quote}

Thus energy-efficiency as quality is a relative quality: Given two
different architectural solutions A and B providing the otherwise same
behavior, measure the energy consumption of each per transaction (or
per second, or over the total running time), and compare to find the
one, A or B, that delivers its services with the least amount of energy
spent.

Therefore teaching in energy consumption will typically reflect both a
theoretical as well as practical/operational level, balancing topics
such as relevant research literature, architectural and programming
techniques for achieving energy efficiency, establishing laboratory
settings for measuring energy consumption and data analysis.

The three courses outlined below demonstrate approaches with a varying
balance of these teaching elements.


\section{Elective Course: Energy Consumption}
\label{section:ruc}

The course is an elective course of 5 ECTS at Masters Education
Computer Science and Masters Education Digital Transformation at
Roskilde University. In Fall 2023, there were 11 3rd semester students. 

%
%
The learning outcomes of the course
\begin{itemize}
    \item know about current research trends related to energy consumption across various relevant fields and disciplines
    \item know about the parameters affecting the energy consumption of software,
    \item demonstrates knowledge about the data analysis of energy estimations/measurements,
    \item demonstrates knowledge about methods for evaluating energy efficiency and energy consumption of IT systems,
    \item has skills in using tools for evaluating energy efficiency and energy consumption of IT systems, and
    \item has competencies in analyzing energy-related issues in IT system
\end{itemize}
The examination is in the form of individual 20 minutes oral exam based on a written product and the assessment will be based on the extent to which the student satisfies the learning outcomes. 

The course has an intended total workload of 135 hours: 30 confrontation hours divided into 10 seminars, 70 hours of preparation and exercises/individual work, and 35 hours for the exam and preparation hereof.

\subsection{Course Design}

The course delves into topics such as experimental design, factors
influencing energy consumption, and two paradigms for measuring and
estimating the efficiency of IT systems. These paradigms include
software-based energy-estimation models (e.g., Intel's RAPL) and
external measurements (e.g., utilizing Siglent SPD3303X-E Programmable
Power Supply). Additionally, the course covers statistical analysis
methods for understanding energy consumption.

\begin{figure}
\begin{vtimeline}[description={text width=6cm},
        row sep=3ex,
        use timeline header,
        timeline title={Elective Course: Energy Consumption (weekly)}
        ]
        1 & Motivation and Practical introduction to Energy Estimation in IT using RAPL. \endlr
        2 & Experimental design, Energy Parameters, and Reproducible vs Replicability\endlr
        3 & Statistical Analysis, and Repeatability and Automation\endlr
        4 & 1st Assignment: Estimation using Software models (RAPL) \endlr
        5& Assignment Feedback \& Using external measurement techniques for power and time. \endlr
        6 & Tuning Experiments and Converting Power measurements to Energy Consumption \endlr
        7 & 2nd Assignment: Measuring using external measurement (Siglent)\endlr
        8 & Assignment Feedback \& Mini-project\endlr
        9 & Mini-project \endlr
        10 & Mini-project\endlr
\end{vtimeline}
\caption{Weekly timeline for the Elective Course: Energy Consumption }
\label{timeline:ECEC}
\end{figure}

To obtain experience using both the software-based energy-estimation models and the external measurements, the participants have to solve two assignments using each of the methods. The assignments are loosely bounded in form of a vague research question, which they have to refine; this ensures that the participants practice how to create hypothesis that can be tested and that they can focus on parameters of their interest. To ensure that they experience the differences arising from the choice of methodology, the participants are only allowed to variate the methodology. In addition, to assess their understanding of the advantages and disadvantages of each of the methods we incorporate a mini-project where they freely choose the research question and  reason about their choice of experimental design.
The course is structured such that there are seminars dedicated to supervision on assignments and mini-project. The weekly timeline for the course can be seen in Figure~\ref{timeline:ECEC}.

The written product for the exam are individual portfolios with their mini-project and improved versions of the assignments.


\subsection{Results}
\subsubsection{Questionnaire}

The final evaluation of the course contained three questions relating
to the topic of energy-efficient . These were formulated as statements
to be evaluation on a Likert scale from \emph{Highly agree} over
\emph{Neutral} to \emph{Highly disagree}. The questions were
\label{sec:maja-questionaire}
\begin{enumerate}
  \item[Q1] \emph{The course has increased my awareness that my
    decisions regarding technology choices, implementation styles, and
    hardware have a direct impact on the energy consumption of IT
    systems.}
  
  \item[Q2] \emph{The course's focus on energy consumption and
    energy-efficiency will be part of, and inspire, my future work, in
    order to lower energy consumption through choices of appropriate
    tactics to reduce energy (use of theory andl earnings from the
    course).}
    
    \item[Q3] \emph{I expect that I will try to make concrete
      measurements and experiments with energy-efficiency in future
      projects (use of practical experimental techniques).}
\end{enumerate}

Out of the 11 participants in the course, 8 answered the
questionnaire. The results are summarized in Table~\ref{table:likert-RUcourse}.

\begin{table}
  \centering
  \caption{Questionnaire answers.}
  \label{table:likert-RUcourse}
  \begin{tabular}{l|rrrrr}
    \hline
    Q & H. Agree & Agree & Neutral & Disagree & H. Disagree \\
    \hline
    Q1 & 6 (75\%) & 2 (25\%) & 0 & 0 & 0 \\
    Q2 & 4 (50\%) & 1 (13\%) & 3 (38\%) & 0 & 0 \\
    Q3 & 3 (38\%) & 1 (13\%) & 2 (25\%) & 2 (25\%) & 0 \\
  \hline
\end{tabular}
\end{table}

The majority of participants (75\%) highly agree that the course has increased their awareness of the impact of technology choices on energy consumption. Only 50\% of participants agree that the course's focus on energy consumption will inspire their future work to lower energy consumption.
The responses are more evenly distributed, with 38\% agreeing that they will make concrete measurements and experiments with energy efficiency in future projects.

\subsubsection{Reflections in Reports}
In the report template used for handing in the assignments and mini-project, there was a specific section on ``Reflections'' where the students could volunteer to write about their experiences. This is widely interpreted by the students. 

In the following we provide the themes identified in the answers and discuss the impact and give quoting examples. The students responses cover 4 themes:
\begin{enumerate}
\item Surprises in Energy Consumption and Code Optimization\label{Course:Finding1}
\item Application of Sustainable Coding Practices\label{Course:Finding2}
\item Challenges in Experimentation and Measurement Techniques\label{Course:Finding3}
\item Critical Thinking and Methodological Improvements\label{Course:Finding5}
\end{enumerate}
The first most common theme reveals that the students came across unexpected results in the relationship between code changes and energy consumption, e.g.,  \emph{``After having conducted the experiment it’s obvious to me that there is not always a significant change of energy consumption when there is an addition of code to a program''}, and  \emph{``I expected that optimized versions of the algorithm to have a positive impact on the energy consumption but it appeared it was the opposite''}.
This theme highlights the gap between theory and practice and the current importance of empirical testing.
The second theme highlights an interest in sustainable coding practices, especially in the context of energy consumption, e.g., \emph{``Arriving on the conclusion that I in my own projects should delete debugging print statements before deployment, I searched for automated solutions.''} This theme reflects a growing awareness of environmental impacts in the field of software engineering.
The third theme concerns the level of knowledge needed for evaluating energy consumption.
The students express frustration about the technical difficulties with hardware and software tools, such as Linux environments and handling hardware like Raspberry Pi, e.g., \emph{``Working with hardware-based measurement was significantly more cumbersome than software-based ones''},  \emph{``It was, as usual, cumbersome to deal with both Linux's sudo (especially when trying to manipulate files from Windows) and with a GUI-less environment, though I also understand the necessity of it''}, and \emph{``One of the most cumbersome part of the project was to ensure internal validity and consider all these steps so the results are not corrupted.''}. This suggests a need for a more easy-approachable technique or setup. 
The last theme identifies suggestions for improving research methodology and considerations for future experiments. The students reflect on their results critically, e.g.,  \emph{``After having conducted the experiment, I found that the Kilo watt hour usage of the two programs to be ~10 which seems like quite a lot which makes me believe that there is something in the experimental setup which are not completely true to nature as this seems to be much more than would be expected of a normal computer.''} and they consider future improvements, e.g, \emph{``To make this a better experiment we will then need to run more experiments and a test with the screen on and off''}. The theme demonstrates the students' engagement in critical thinking and their ability to reflect on their learning process.

In conclusion, the reflections align well with the answers to the questionnaire and, in addition, provide some possible explanations into Q3, namely the cumbersome techniques and the amount of effort required to obtain reliable energy evaluations.


\section{Software Architecture in Practice}
\label{section:saip}

The course context is a 15 ECTS module, divided in three courses of 5
ECTS each, that is taught at Department of Computer Science,
Aarhus University,
as part of
Danish
lifelong education effort. Each of the three courses are
quarter length, lasting 7 weeks each. The three courses are a
progression, building upon the previous course, and forms a whole. The
course is at Master's level, and students are part-time students,
typically between 30-60 years old, and employed as programmers and
software architects in the software and electronic industry. Thus, the
focus is \emph{theory and methods in order to improve the practice of
  doing software architecture and engineering}.

The learning goals, teaching plan and exercises are about software
architecture at large, so the experiences reported below stems from
one week of teaching on energy-efficiency and one major written
exercise that is part of the second course's exam.

The exercise defines the \emph{Learning Goal: To experiment with
  energy-efficiency in the TeleMed~\cite{repo:telemed} system by
  setting up a concrete two machine lab and set up JMeter work load
  models for conducting energy experiments, do systematic measurements
  of an A/B variant of the TeleMed architecture following guidelines
  by Cruz~\cite{cruz:2021}, do statistical analysis and reporting of
  the findings and conclude.} The actual A/B variants are outlined
below.

\subsection{Theory}

There is a lot of research and engineering on the topic of
energy-efficiency but the field is still rather immature and still
lacks an overall framework and ontology. Furthermore, there are many
concrete practices and tactics for making software consume less
energy, but the resources are often dedicated to specific areas
(notably app development for smartphones). Inspired by the tactics
frameworks of \cite{bass:2021}, one of the authors has compiled a
framework for energy saving tactics: \emph{The Green Architecture
  Framework} (GAF)~\cite{gaf}. This framework outlines more than 20
tactics\footnote{A tactic is a design decision that influences a given
  architectural quality attribute~\cite{bass:2021}.} organized in
seven categories. Examples of tactics that (may) lower energy
consumption are \emph{Use Batch Method Pattern}: transfer many items
over a network in one large network package instead of a single item
each in one small network package; \emph{Use Low-footprint Data
  Formats}: avoid overhead in verbose dataformats (like prefer JSON or
ProtoBuf over XML); \emph{Use Efficient Databases}: Use key-value
stores (like Redis) rather than a SQL database, if there is no need
for an advanced query language; or \emph{Lower Fidelity of
  Video/Images}: downscaled images have smaller footprints which
consumes less energy in transit and when decompressing. For full
details, consult~\cite{gaf}.

Thus, our theoretical presentation was rooted in GAF, supplemented
with presentations of measuring techniques~\cite{cruz:2021} and
statistics, as well as presented a series of our own experiments on
database choices, logging, choice of programming language,
etc. Further detail can be found in~\cite{christensen:saip-week12}.

\subsection{Exercise}
The course exercises adopt a storytelling approach~\cite{christensen:2009}
and thus all are rooted in the same case, TeleMed~\cite{repo:telemed},
which is a prototype tele-medical application: a client-server system
to allow patients to measure and upload blood pressure measurements
and allow general practitioners to review them.  The concrete exercise
focuses on measuring energy and experimenting with energy-efficiency
in this TeleMed system.

The concrete exercise consists of two (rather work intensive)
prerequisites and three concrete experiments.

The prerequisites are to A) setup a lab consisting of two physical
machines, one hosting the TeleMed system and one hosting a load
generator (Apache JMeter); and B) refine a work load profile from a
previous exercise, to generate high load on the TeleMed system.

The three concrete exercises were based upon the hypotheses that A)
Switching the TeleMed database from MongoDB to Redis will save energy;
B) Turning logging off in the TeleMed system will save energy; and C)
a free pick between a set of energy relevant exercises (change of
storage format, running as Docker container, a.o.)

Their reporting was supported by providing a writing template for
their work. Note that as per the definition of energy-efficiency, all
exercises involve two experiments, A and B, that must be reported and
compared.

The framework of Cruz~\cite{cruz:2021} was required for controlling
the experiments and for statistical analysis of results. Typically the
latter involved violin-plots and Welch t-tests.


\subsection{Results}

There are two sources for students' experience with the exercise: a
questionnaire with three questions as part of the final course
evaluation; and a required section in their hand-in under the headline ``Reflections on Energy Experiments''.

\subsubsection{Questionnaire}

The final evaluation of the course contained three questions relating
to the topic of energy-efficient architecture. These were formulated
as statements to be evaluation on a Likert scale from \emph{Highly
  agree} over \emph{Neutral} to \emph{Highly disagree}. The questions
were similar to those reported in Section~\ref{sec:maja-questionaire}.

      

Out of the 19 participants in the course, 14 answered the
questionnaire. The results are summarized in Table~\ref{table:likert}.

\begin{table}
  \centering
  \caption{Questionnaire answers.}
  \label{table:likert}
  \begin{tabular}{l|rrrrr}
    \hline
    Q & H. Agree & Agree & Neutral & Disagree & H. Disagree \\
    \hline
    Q1 & 6 (43\%) & 6 (43\%) & 2 (14\%) & 0 & 0 \\
    Q2 & 1 (7\%) & 2 (14\%) & 8 (57\%) & 2 (14\%) & 0 \\
    Q3 & 0 & 2 (14\%) & 6 (43\%) & 5 (36\%) & 1 (7\%) \\
  \hline
\end{tabular}
\end{table}

The conclusion on the questionnaire results is that student's work
with the theory as well as their concrete experiments have greatly
increased awareness of energy-efficiency; that their experiences will
to some extend influence their future work in industry; but also that
it is perceived less likely that they will be investing in making
actual experiments. Given the industry's rather stressful work
environment with tight deadlines, this is perhaps not surprising.

\subsubsection{Reflections in Reports}

In the report template to be used for handing in their mandatory
exercise, there was a specific section with the headline ``Reflections
on Energy Experiments''. There were eight groups and all but a single
group wrote about their experiences. Broadly, they report in two
categories: \emph{energy savings-} and \emph{process-} experiences.

Regarding the actual results of experiments, multiple groups report
some amazement on just how much energy can actually be saved, for
instance by reducing logging. Obviously, the \emph{awareness} on
energy consumption of software has been awakened.

Process issues were discussed along more dimensions by the
groups. First, most reported surprise and ``aha'' moments during their
work, and one group reported it as a ``fun and challenging
exercise''. However, the most pervasive reflection was about how hard
and labour intensive it is to set up a controlled environment, conduct
the experiments many times, and do the proper statistical
analysis. Several noted the high important of \emph{automation} in the
process to lower the workload. And, as our students are working in the
software industry in their day job, many noted that the investment in
time ``is not justified in an industrial context''. One group noted
that energy-efficiency as a result must be considered already in the
architecting phase.

In conclusion, these reflections align well with those of the
questionnaire: the exercise has increased awareness on
energy-efficiency but controlled experiments are time consuming and
experiments deemed too cumbersome.  Perhaps our changing climate will
change that perception in the future\ldots

\section{Specialization in Energy Aware Programming}

\label{section:aau}

Since the autumn 2020 we have, as part of the MSc programme in
software engineering at Aalborg University, offered a specialization
year in Energy Aware Programming as part of the programming technology
specialization. This specialization has been chosen by 6 students in
2020, 12 in 2021, 7 in 2022 and 5 in 2023.  The specialization is the
major part of the final year of the programme and consists of a 5 ECTS
course and a 20 ECTS project in the autumn term and a 30 ECTS project
(the master thesis project) in the spring term. The project work is
done in groups and follows the Problem Based Learning principles.

In the specialization course the students learn how to read, analyze
and present scientific papers. The curriculum states that the students
will gain in-depth insight into central topics within research in
programming technology and will be able to provide a clear and
comprehensible presentation and discussion of articles' central
topics, including their premises, issue(s), theory, methods, results
and conclusions. During the course each student will present three
scientific papers and read up to 30 papers. The first paper will be
assigned by the teachers and the following two papers the students
will select guided by the teachers. Often these papers are relevant to
their group project work, which ensures that each year there is a new
reading list. The course exam is an oral exam where the student is
assigned a paper to present and relate to the body of knowledge
accumulated during the course. The student has one week to prepare for
the exam.

The 2022 course involved reading, analyzing, and presenting the following papers:
\cite{pereira2017energy,cruz2019energy,lima2016haskell,koedijk2022finding,zhu2022learning,hamizi2021meta,georgiou2020energy,von2018measuring,cuadrado2017comparative,mancebo2021process,bokhari2020towards,fahad2019comparative,hansen2020benchmarking,bruce2018approximate,sestoft2013microbenchmarks,de2022webassembly,ournani2021tales,hackenberg2013power,mancebo2021feetings,ournani2021evaluating,hindle2015green,khan2018rapl,csanlialp2022energy,rajan2016study}

As an example, one student presented the paper \cite{mancebo2021feetings} which presents a Framework for Energy Efficiency Testing to Improve eNvironmental Goals of the Software (FEETINGS) comprising an ontology, a process that helps researchers to evaluate the energy efficiency of the software and two technologies for measuring and statistical analysis of results. The student related the paper to several papers introducing processes and/or technologies for testing energy consumption of software \cite{bokhari2020towards,von2018measuring}. The student’s own opinion of the paper was that the ontology and process were very useful, although the process had a few unnecessary steps and in other cases lacked details. However, the technologies were perceived as outdated. The students went on to use the ontology and process, with minor modification, in the semester project, replacing the technology component with their own based on RAPL.

The ability to dissect, present and discuss scientific papers is clearly a needed skill for MSc students wanting to enter the PhD programme and pursue an academic career. It is, however, also a very useful skill for students wanting to pursue a career in the software industry where scientific innovations are turned into useful systems, especially the ability to assess the practical usefulness of proposed methods, solutions and results, as well as the ability to assess the scalability and transferability of results. This is in particular so for the topic energy aware programming, as this subject is not yet a well developed discipline.
As part of a general quality assurance framework, the course is evaluated by the students. The students are asked the following three questions, with answers from the 2022 run of the course added: 

\begin{enumerate}
\item Point out things that worked well and come up with ideas on how they could work even better

\begin{itemize}
    \item Quite enjoyed everyone coming in to present and analyse papers
\item Learning how to read papers is a very valuable skill and necessary in the report/thesis
\item Felt like good interaction with everyone in the class, even if students were sometimes hesitant to speak up
\item Excellent interaction with professors, they've done very well
\end{itemize}

\item Point out things that worked less well and come up with ideas on how they could be improved

\begin{itemize}
    \item 

Having to both present and review on the same day requires a lot of preparation. You should probably spread it out then
as much as possible
\end{itemize}

\item Possible supplementary comments

\begin{itemize}
    \item A lot of students had to cancel/reschedule their presentations, or just sometimes didn't show up. This isn't the fault of those
running the course, but it is at least worth mentioning because it's bothersome to us, the students
\item The course seemed like it had enough extra time planned to handle this, however, which is good
\end{itemize}
\end{enumerate}

Over the years a total of 120 papers have been read and 90 presented. This has given unique insight and overview of the literature in the area of energy aware programming.
First of all it has revealed that this area of research is scattered over a broad spectrum of outlets. There is a huge difference in scientific approach, i.e. how to establish lab conditions for performing energy measurement experiments, which statistical methods to use and even the number of samples needed. Other questions pertain to the nature of benchmarks, e.g. micro- vs. macro- or application benchmarks.
These discussions have been used by the students as inspiration in their projects.

During their specialization, students undertake a 20 ECTS pre-specialization project, typically in groups of up to six students. This project aims to deepen their understanding of a current software engineering or computer science research problem and prepare them for their thesis. 
They document their findings in a report or scientific paper.
As part of the project, the students have to survey the
state-of-the-art, clearly taking outset in the articles from the
specialization course. The students usually carry out hands-on
experimental work inspired by the knowledge they gain. So far six
pre-specializations projects have been completed. An example of such a
project is entitled "The Influence of Programming Paradigms on Energy
Consumption" where students investigated how the programming style
would affect energy consumption by (re)writing a number of benchmarks
from CLBG in C\# in an imperative, object oriented and a functional
style. Surprisingly this work shows that sometimes a pure functional
approach can be the most energy efficient approach. Another example is
a project entitled “Energy Benchmarking With Doom: Utilizing video
game source ports for macrobenchmarking”, where the students showed
that the C\# version of Doom is as energy efficient as the C and C++
versions.

The final six months of the MSc programme are spent on a master thesis
project done individually or in groups of up to three students. The
purpose of this project is that the students can formulate, analyze
and contribute to solving a current research problem in software engineering independently, systematically and critically through the
application of scientific theory and method. The students will
document in-depth knowledge and overview of a current problem and its
possible solutions.  So far 11 master thesis projects have been
completed. An example of a master thesis project is entitled "Energy
Consumption of Software Architectures - A comparison between
microservice- and monolithic architectures in C\# and Java" where the
students investigated four different versions of the Pet Store web
application benchmark. The main result shows that a monolithic
architecture was the most energy efficient, at least to the level of
scalability possible in their set up \cite{Gregersen2022}.

The presentations and discussions in the specialization course and the supervision of pre-specialization projects and master thesis projects have given the teachers insight into the scientific literature and possible solutions to many open questions. This insight has been distilled in discussion with Author 1 and Author 2 and (hopefully) helped them in designing their courses.

\section{Discussion}

The perhaps most important lesson learned and the \emph{key takeaway}
from our experiences is that of \emph{increasing awareness about
  energy consumption in the student population.} Generally, students
report that it was an \emph{eye-opening} experience to work with
energy in relation to software development: Most students had never
thought that the way they program their systems influences the energy
consumption to such a far extent!

Regarding student perception of energy consumption it is interesting
to note the difference between the student populations in the courses
presented in Section~\ref{section:ruc} and~\ref{section:saip}. Table~1 and
Table~2 represent answers to identical questions but Table 1 are
answers from ordinary (young/industrial inexperienced) master students
while Table 2 are answers from part-time (older/experienced)
students. The experienced students are less likely to answer ``highly
agree''---especially in Q3 where 51\% young students think they will
conduct energy experiments whereas only 14\% of the experienced
developers think that. While some may perhaps be attributed to age,
we also hypothesize that it is due to the more experienced developers
knowing that there may not be management support to justify the cost,
something that may change as customers demand accountability,
e.g. some Danish authorities now require software vendors to account
for sustainability in their development practice.

We acknowledge that the small sample size of the two questionnaires is
a threat to validity. Both courses had a small student population, and
we did make answering the questionnaires voluntary. Thus, we of course
cannot make any general claims. Still, we find that the
``eye-opening'' experience that we have ourselves had, and often
reported anecdotal by the students, is supported from the output of
the two questionnaires.

Across the courses, students report the high workload associated with
reproducible data collection from energy experiments. Guidelines like
that of Cruz~\cite{cruz:2021} help in a good process, but
never-the-less many hours must be spent on setting up a correct lab,
repeating experiments, and doing statistical analysis. We find that
more research on recommendations and best-of-breed practices is
required to support teachers in setting up quality environments, labs,
and automating measurements. This may be especially important if the
subject is taught as part of a larger software engineering course or
spread across several courses in a software engineering curricula.

This aligns with the insight that our students are rather
inexperienced with experimental studies: the need for a controlled
environment, repeated experiments, and statistical analysis. Thus,
there is an initial learning curve to be climbed in motivating the
need for a large set of experiments, attention to detail in the
environment, and the concrete statistical methods.

Many research papers report on energy measurements and lab settings
that emphasize solely on reproducibility to the extent that practical
applicability of the results can be questioned. Examples are
experiments where hyper-threading and turbo boost is turned off, or
execution is pinned to a single thread/core. While it may lower error
sources it also removes the ability of modern hardware and operating
systems to optimize and thus execute energy efficient. We suggest that
teachers include this consideration in their teaching.

Sometimes students lack intuition about their results - they do not
always have a back-of-an-envelope calculation capability to judge
their results and may therefore draw wrong conclusions or overstated
conclusions.  We encourage our students to go beyond their results,
i.e. when they measure that code for program P compiled with compiler
A produces code that is X\% more energy efficient than code for P
produced by compiler B, and we encourage them to analyse the machine
instructions generated to assess if the difference seems plausible and
similarly when comparing algorithms, program patterns or library
calls. Far too many scientific papers fail to do such analysis and
often leave the reader wondering about such explanations.

\section{Conclusion}

In this paper, we have outlined three different approaches for
teaching sustainability, green computing, and energy-efficient
software development.

Our main contribution is presenting viable teaching techniques that
address the issues pertaining to software development and its energy
consumption, with a spectrum from literature review to experimental
work, as well as lessons learned. We have identified a number of
issues that influence the learning experience and must thus be
factored in when designing a course. Experimental work is time
consuming but the students involved in it generally describe it as an
eye-opening experience as the energy savings can be quite big for even
minor changes.

Perhaps the most important message of teaching energy aware software
development is the \emph{awareness} that we as software developers
also can contribute in combating the climate breakdown, and as such
it needs our attention.

\bibliographystyle{ACM-Reference-Format}
\bibliography{bib}
\end{document}